\documentclass[proceedings, preprint]{rmaa}

\usepackage{paralist}

\usepackage{psfrag,color}

\SetYear{2012}
\SetConfTitle{Adela: Astronom\'ia Din\'amica en Latinoam\'erica}

\title{Circumnuclear star-forming regions in early type spiral galaxies: dynamical masses} 

\author{
  G. F. H\"agele,\altaffilmark{1,2}
  M. V. Cardaci,\altaffilmark{1,2}
  G. L. Bosch,\altaffilmark{1,2}
  A. I. D\'{\i}az,\altaffilmark{3} 
  E. Terlevich,\altaffilmark{4}
  and R. Terlevich\altaffilmark{4}}

\altaffiltext{1}{Facultad de Cs.\ Astron\'omicas y Geof\'isicas, Universidad
  Nacional de La Plata, Paseo del Bosque s/n, 1900 La Plata, Argentina}

\altaffiltext{2}{IALP-Conicet, Paseo del Bosque s/n, 1900 La Plata, Argentina.}

\altaffiltext{3}{Departamento de F\'{\i}sica Te\'orica, M\'odulo 15, Universidad
  Aut\'onoma de Madrid, 28049 Madrid, Spain}
\altaffiltext{4}{Instituto Nacional de Astrof\'isica, \'Optica y Electr\'onica,
  Tonantzintla, Apdo. Postal 51, 72000 Puebla, M\'exico}

\shortauthor{H\"agele et al.}
\shorttitle{RevMexAA(SC) Demo Document}

\listofauthors{G. F. H\"agele, M. V. Cardaci, G. L. Bosch, A. I. D\'{\i}az,
  E. Terlevich, \&R. Terlevich}
\indexauthor{G. F. H\"agele}
\indexauthor{M. V. Cardaci}
\indexauthor{G. L. Bosch}
\indexauthor{A. I. D\'{\i}az}
\indexauthor{E. Terlevich}
\indexauthor{R. Terlevich}

\abstract{
We present the measurements of
gas and stellar velocity dispersions in 17 circumnuclear
star-forming regions (CNSFRs) and the nuclei of three barred spiral galaxies:
NGC\,2903, NGC\,3310 and NGC\,3351 from high dispersion spectra. The stellar
dispersions have been obtained from the Ca{\sc ii} triplet (CaT) lines at
$\lambda\lambda$\,8494, 8542, 8662\,\AA, while the gas velocity dispersions
have been measured by Gaussian fits to the H$\beta$\,$\lambda$\,4861\,\AA\ and
to the [O{\sc iii}]\,$\lambda$\,5007\,\AA\ lines.  

The CNSFRs, with sizes of about 100 to 150\,pc in diameter, are seen to be
composed of several individual star clusters with sizes between 1.5 and
6.2\,pc on {\it Hubble Space Telescope} (HST) images. Using the stellar
velocity dispersions, we have 
derived dynamical masses for the entire star-forming complexes and for the
individual star clusters. Values of the stellar velocity dispersions are
between 31 and 73\,km\,s$^{-1}$. Dynamical masses for the whole CNSFRs are
between 4.9\,$\times$\,10$^6$ and 1.9\,$\times$\,10$^8$\,M$_\odot$ and
between 1.4\,$\times$\,10$^6$ and 1.1\,$\times$\,10$^7$\,M$_\odot$ for the
individual star clusters. 

We have found indications for the presence of two different kinematical
components in the ionized gas of the regions. The narrow component of the
two-component Gaussian fits 
seem to have a relatively constant value for all the studied CNSFRs, with 
estimated values close to 25\,km\,s$^{-1}$. This narrow component could be
identified with ionized gas in a rotating disc, while the stars and the
fraction of the gas (responsible for the broad component) related to the
star-forming regions would be mostly supported by dynamical pressure. 
}

\resumen{Presentamos medidas de la dispersi\'on de velocidades en 17 regiones
  de formaci\'on estelar circunnucleares (CNSFRs) y los n\'ucleos de tres
  galaxias espirales barradas: NGC\,2903, NGC\,3310 y NGC\,3351 a partir de
  espectros de alta dispersi\'on. Las dispersiones estelares han sido
  obtenidas de las l\'ineas del triplete de CaII (CaT) a
  $\lambda\lambda$\,8494, 8542, 8662\,\AA, mientras que las dispersiones de
  velocidades del gas han sido medidas mediante ajustes gaussianos a las
  l\'ineas de H$\beta$\,$\lambda$\,4861\,\AA\ y de
  [OIII]\,$\lambda$\,5007\,\AA. 
  
  Las CNSFRs, con tama\~nos de alrededor de 100 a 150pc en di\'ametro, parecen
  estar compuestas por varios c\'umulos estelares individuales con tama\~nos
  entre 1.5 y 6.2pc medidos sobre imagenes del Telescopio Espacial Hubble
  (HST). Utilizando las dispersiones de velocidades estelares, hemos derivado
  las masas din\'amicas para los complejos de formaci\'on estelar completos y
  para los c\'umulos estelares individuales. Los valores de la dispersi\'on de
  velocidades estelares est\'an entre 31 y 73\,km\,s$^{-1}$. Las masas
  din\'amicas para 
  las CNSFRs completas est\'an entre 4.9\,$\times$\,10$^6$ y
  1.9\,$\times$\,10$^8$\,M$_\odot$ y entre 1.4\,$\times$\,10$^6$ y
  1.1\,$\times$\,10$^7$\,M$_\odot$ para los c\'umulos estelares individuales. 

Hemos encontrado indicaciones de la presencia de dos componentes cinem\'aticas
diferentes en el gas ionizado de las regiones. La componente estrecha
proporcionada por el ajuste de dos-componentes gausianas, parece tener un
valor relativamente constante para todas las CNSFRs estudiadas, con los
valores estimados muy pr\'oximos a  25\,km\,s$^{-1}$. Esta componente estrecha
podr\'ia 
identificarse con gas ionizado en un disco rotante, mientras que las estrellas
y la fracci\'on de gas (responsable de la componente ancha) relacionadas con
las regiones de formaci\'on estelar, estar\'ian mayormente soportadas por
presi\'on din\'amica.  
}

\addkeyword{H~II regions}
\addkeyword{galaxies: kinematics and dynamics}
\addkeyword{galaxies: starburst}
\addkeyword{galaxies: star clusters}

\begin{document}
\maketitle

\section{Introduction}
\label{sec:intro}

Gas content, masses, bar structure, and dynamical environment can strongly
influence the large-scale star formation rate (SFR) along the Hubble sequence
\citep{1998ARA&A..36..189K}. The variation of young stellar content and star
formation activity is one of the most conspicuous characteristic along this
sequence, and this variation in the young stellar population is part of the
basis of the morphological classification made by
\citet{1926ApJ....64..321H}. The trend in SFRs and star formation histories
along the Hubble sequence was confirmed from evolutionary synthesis models of
galaxy colours by \citet{1968ApJ...151..547T,1972A&A....20..383T} and
\citet{1973ApJ...179..427S}. Later, the importance of the star formation
bursting mode in the evolution of low-mass galaxies and interacting systems
was studied by \citet{1976PhDT.........6B,1977ApJ...217..928H} and
\citet{1978ApJ...219...46L}. Due to their different average SFR, the
integrated spectra of galaxies vary considerably along the Hubble sequence.

The gas flows in disc of spiral galaxies can be strongly perturbed by the
presence of bars, although the total disc SFR does not appear to be
significantly affected by them \citep{1998ARA&A..36..189K}. These perturbations
of the gas flow trigger nuclear star formation in the bulges of some barred
spiral galaxies. These structures around the nuclei of some
spiral galaxies present higher than usual star 
formation rates and are frequently arranged in a ring pattern with
a diameter of about 1\,Kpc. At optical wavelengths, these circumnuclear
star-forming regions (CNSFRs) are easily observable rings.
Although CNSFRs are very luminous, not much is known about their 
kinematics or dynamics for both the ionized gas and the stars. In fact, the
most poorly known property of star forming clusters in galaxies is their
mass.

\section{Observations and data reduction}
\label{obs-kine}

\subsection{Observations}

High resolution blue and far-red spectra were acquired as part of an observing
run in 2000. They were obtained simultaneously using the blue and red arms of
the Intermediate dispersion Spectrograph and Imaging System (ISIS) on the
4.2-m William Herschel Telescope (WHT) of the Isaac Newton Group (ING) at the
Roque de los Muchachos Observatory on the Spanish island of La Palma. The
H2400B and R1200R gratings were used to cover 
the wavelength ranges from 4779 to 5199\,\AA\ ($\lambda_c$\,=\,4989\,\AA) in
the blue and  from 8363 to 8763\,\AA\ ($\lambda_c$\,=\,8563\,\AA) in the red
with spectral dispersions  of 0.21 and 0.39 \AA\ per pixel, equivalent to a
spectral resolution (R\,=\,$\lambda$\,/\,$\Delta\lambda$) of $\sim$\,23800 and
$\sim$\,22000, respectively, and 
providing a comparable velocity resolution of about 13\,km\,s$ ^{-1}$. The CCD
detectors EEV12 and TEK4 were used for the blue and red arms with a factor of
2 binning in both the ``x" and ``y" directions in the blue with spatial
resolutions of 0.38 and 0.36 \,arcsec\,px$^{-1}$ for the blue and red
configurations respectively. A slit width of 1\,arcsec was used which,
combined with the spectral dispersions, yielded spectral resolutions of about
0.4 and 0.7\,\AA\ FWHM in the blue and the red, respectively, measured on the
sky lines.

In the cases of NGC\,2903 (H\"agele et al.\ 2009) and NGC\,3310 (H\"agele et
al.\ 2010) two different slit positions were
chosen in each case to observe 4 and 8 CNSFRs, respectively, which we have  
labelled S1 and S2 for each galaxy. Besides, for NGC\,3310 we observed the
conspicuous Jumbo region. The name of 
this region, Jumbo, comes from the fact that it is 10 times more luminous than
30 Dor \citep{1984ApJ...284..557T}. For the third galaxy, NGC\,3351 (H\"agele
et al.\ 2007), three different slit positions were chosen to observe 5
CNSFRs. In all the cases one of the slits passes across the nucleus.

Several bias and sky flat field frames were taken at the beginning and the end
of each night in both arms. In addition, two lamp flat field and one
calibration lamp exposure per each telescope position were performed. The
calibration lamp used was CuNe+CuAr.

We have also downloaded two astrometrically and photometrically calibrated
broad-band images of the central part of each observed galaxy from the
Multimission Archive at Space  
Telescope\footnote{http://archive.stsci.edu/hst/wfpc2}. The images were taken
through the F606W (wide V)  and the F160W (H) filters with the
Wide Field and Planetary Camera 2 (WFPC2; PC1) and the Near-Infrared Camera
and Multi-Object Spectrometer (NICMOS) 2 (NIC2; for NGC\,3310) and 3
(NIC3; for NGC\,2903 and NGC\,3351), both cameras on-board the HST. In the
case of NGC\,3310 we also download the F658N narrow band image (equivalent to
H$\alpha$ filter at the NGC\,3310 redshift) taken with the Advanced Camera for
Surveys (ACS) of the HST. All the measured sizes of the circumnuclear regions
are defined using these high resolution images.

\subsection{Data reduction}

The data was processed and analyzed using IRAF\footnote{IRAF: the Image
  Reduction and Analysis Facility is distributed by the National Optical
  Astronomy Observatories, which is operated by the Association of
  Universities for Research in Astronomy, Inc. (AURA) under cooperative
  agreement with the National Science Foundation (NSF).} routines in the usual
manner. The procedure includes the removal of cosmic rays, bias subtraction,
division by a normalized flat field and wavelength calibration. Wavelength
fits were performed using 20-25 arc lines in the blue and 10-15 lines in the
far-red by a polynomial of second to third order. These fits have been done at
50 and 60 locations along the slit in the blue and far-red, respectively, and
they have yielded rms residuals between $\sim$0.1 and $\sim$0.2\,px. 
We have not corrected the spectra for atmospheric extinction or performed any
flux calibration, since our purpose was to measure radial velocities and
velocity dispersions.

In addition to the galaxy frames, observations of 11 template velocity stars
were made to provide
good stellar reference frames in the same system as the galaxy spectra for the
kinematic analysis in the far-red. They are late-type giant and supergiant
stars which have strong CaT features.


\section{Summary and Conclusions}

We present the measurements of gas and stellar velocity dispersions in 17
circumnuclear star-forming regions (CNSFRs) and the nuclei of three barred
spiral galaxies: NGC\,2903 (H\"agele et al.\ 2009), NGC\,3310 (H\"agele et
al.\ 2010) and NGC\,3351 (H\"agele et al.\ 2007) from high dispersion
spectra. The stellar 
dispersions have been obtained from the Ca{\sc ii} triplet (CaT) lines at
$\lambda\lambda$\,8494, 8542, 8662\,\AA that originates in the atmospheres of
red giant and supergiant stars belonging to the underlying stellar population
of the clusters, while the gas velocity dispersions 
have been measured by Gaussian fits to the H$\beta$\,$\lambda$\,4861\,\AA\ and
to the [O{\sc iii}]\,$\lambda$\,5007\,\AA\ lines.  

The CNSFRs, with sizes of about 100 to 150\,pc in diameter, are seen to be
composed of several individual star clusters with sizes between 1.5 and
6.2\,pc on {\it Hubble Space Telescope} (HST) images. Stellar velocity
dispersions are between 31 and 73 km/s. For NGC2903 and NGC3351 these values
are about 25 km/s larger than those derived for the gas from the H$\beta$
emission line using a single Gaussian fit. For NGC3310 these
values and those derived for the gas from the H$\beta$ emission line using a
single 
Gaussian fit are in relatively good agreement, with the former being slightly
larger. However, the best Gaussian fits involved two different components for
the gas: a ``broad component'' with a velocity dispersion similar to that
measured for the stars for NGC2903 and NGC3351, and larger by about 20 km/s
for NGC3310, and a ``narrow component'' with velocity dispersions lower than
the stellar one by about 30 km/s. This ``narrow component'' seems to have a
relatively constant value for all the CNSFRs studied in these three galaxies,
with estimated values close to 25 km/s for the two gas emission lines.

Values for the upper limits to the dynamical masses estimated from the stellar
velocity dispersion using the virial theorem for the CNSFRs of NGC2903 are in
the range between 6.4x10$^7$ and 1.9x10$^8$ M$_\odot$ and is 1.1x10$^7$
M$_\odot$ for its nuclear region inside the inner 3.8 pc. In the case of
NGC3310 the masses are in the range between 2.1x10$^7$ and1.4x10$^8$ M$_\odot$
for the CNSFRs and for the nuclear region inside the inner 14.2 pc is
5.3x10$^7$ M$_\odot$. For NGC3351 the dynamical masses are in the range
between 4.9x10$^6$ and 4.5x10$^7$ M$_\odot$ for the CNSFRs and is 3.5x10$^7$
M$_\odot$ for the nuclear region inside the inner 11.3 pc. Then, globaly, the
dynamical masses of the CNSFRs are in the range between 4.9x10$^6$ and
1.9x10$^8$ M$_\odot$. Masses derived from the H$\beta$ velocity dispersion
under the assumption of a single component for the gas would have been
underestimated by factors between approximately 2 to 4. 

The derived masses for the individual clusters are between 1.4x10$^6$ and
1.1x10$^7$ M$_\odot$, between 1.8 and 7.1x10$^6$ M$_\odot$, and between 1.8
and 8.7x10$^6$ M$_\odot$ for NGC2903, NGC3310 and NGC3351, respectively. Then,
globally, the masses of these individual clusters vary between 1.4x10$^6$ and
1.1x10$^7$ M$_\odot$. 
These values are between 4.2 and 33 times the mass
derived for the SSC A in NGC1569 by \citet{1996ApJ...466L..83H}
and larger than
other kinematically derived SSC masses in irregular galaxies
(McCrady et al., 2003; Larsen et al., 2004). 
It must be noted that we have measured the size of
each knot (typically between 3 and 5pc), but the stellar velocity dispersion
corresponds to the integrated CNSFR wider area containing several knots. The
use of these wider size scale velocity dispersion measurements to estimate the
mass of each knot, leads us to overestimate the mass of the individual
clusters, and hence of each CNSFR. We can not be sure though that we are
actually measuring their velocity dispersion and thus prefer to say that our
measurements of $\sigma_\ast$ and hence dynamical masses constitute upper
limits.  

Masses of the ionizing stellar clusters (derived from their H$\alpha$
luminosities under the assumption that the regions are ionization bound and
without taking into account any photon absorption by dust) for the regions of
NGC2903 are between 3.3 and 4.9x10$^6$ M$_\odot$, and is 2.1x10$^5$ for its
nucleus. The values derived in NGC3310 are between 8.7x10$^5$ and 2.1x10$^6$
M$_\odot$ for the star-forming regions, and is 3.5x10$^6$ for the nucleus. For
NGC3351 are between 8.0x10$^5$ and 2.5x10$^6$ M$_\odot$ for the regions, and
is  6.0x10$^5$ M$_\odot$ for its nuclear region. Thus, the masses of the
ionizing stellar cluster studied in these three galaxies vary between
8.0x10$^5$ and 4.9x10$^6$ M$_\odot$. Therefore, the ratio of the ionizing
stellar population to the total dynamical mass is between 0.01 and 0.16. These
values of the masses of the ionizing stellar clusters of the CNSFRs are
comparable to that derived by Gonz\'alez-Delgado et al.\ (1995) for the
circumnuclear region A in NGC7714. 

Derived masses for the ionized gas (also from their H$\alpha$ luminosities)
vary between 6.1x10$^4$ and 1.3x10$^5$ M$_\odot$ for the regions and is
3x10$^3$ M$_\odot$ for the nucleus of NGC2903; between 1.5 and 7.2x10$^5$
M$_\odot$ for the CNSFRs and is 5x10$^3$ M$_\odot$ for the nucleus of NGC3310;
and between 7.0x10$^3$  and 8.7x10$^4$ M$_\odot$ for the CNSFRs of NGC3351,
and is 2x10$^3$ M$_\odot$ for its nucleus. Globally, the masses of the ionized
gas vary between 7.0x10$^3$ and 7.2x10$^5$ M$_\odot$ for the CNSFRs studied in
these three galaxies. These values are also comparable to that derived by
Gonz\'alez-Delgado et al. (1995) for region A in NGC7714. 

We have found indications for the presence of two different kinematical
components in the ionized gas of the regions. The narrow component of the
two-component Gaussian fits seem to have a relatively constant value for all
the studied CNSFRs, with estimated values close to 25\,km\,s$^{-1}$. This
narrow component could be identified with ionized gas in a rotating disc,
while the stars and the fraction of the gas (responsible for the broad
component) related to the star-forming regions would be mostly supported by
dynamical pressure. To disentangle the origin of these two components it will
be necessary to map these regions with higher spectral and spatial resolution
and much better signal-to-noise ratio in particular for the O$^{2+}$ lines.

The observed stellar and [OIII] rotational velocities of NGC2903 are in good
agreement, while the H$\beta$ measurements show shifts similar to those find
between the narrow and the broad components. In the case of NGC3310 the
rotation curve shows a typical S feature, with the presence of some
perturbations, in particular near the location of the Jumbo region. For
NGC3351, the rotation velocities derived for both stars and gas are in
reasonable agreement, although in some cases the gas shows a velocity slightly
different from that of the stars. For the three galaxies, the rotation curve
corresponding to the position going through the centre of the galaxy shows
maximum and minimum values at the position of the circumnuclear ring.

\end{document}